# Internames: a name-to-name principle for the future Internet

Nicola BLEFARI MELAZZI*, Andrea DETTI*, Mayutan Arumaithurai†, K.K. Ramakrishnan˟
* CNIT - Department of Electronic Engineering, University of Rome "Tor Vergata", Rome, Italy.
† University of Goettingen, Germany, ˟ Rutgers University, U.S.A.
blefari@uniroma2.it, andrea.detti@uniroma2.it, arumaithurai@cs.uni-goettingen.de, kkramakrishnan@yahoo.com

*Abstract*—We propose Internames, an architectural framework in which names are used to identify all entities involved in communication: contents, users, devices, logical as well as physical points involved in the communication, and services. By not having a static binding between the name of a communication entity and its current location, we allow entities to be mobile, enable them to be reached by any of a number of basic communication primitives, enable communication to span networks with different technologies and allow for disconnected operation. Furthermore, with the ability to communicate between names, the communication path can be dynamically bound to any of a number of end-points, and the end-points themselves could change as needed. A key benefit of our architecture is its ability to accommodate gradual migration from the current IP infrastructure to a future that may be a ubiquitous Information Centric Network. Basic building blocks of Internames are: i) a name-based Application Programming Interface; ii) a separation of identifiers (names) and locators; iii) a powerful Name Resolution Service (NRS) that dynamically maps names to locators, as a function of time/location/context/service; iv) a built-in capacity of evolution, allowing a transparent migration from current networks and the ability to include as particular cases current specific architectures. To achieve this vision, shared by many other researchers, we exploit and expand on Information Centric Networking principles, extending ICN functionality beyond content retrieval, easing send-to-name and push services, and allowing to use names also to route data in the return path. A key role in this architecture is played by the NRS, which allows for the co-existence of multiple network "realms", including current IP and non-IP networks, glued together by a name-to-name overarching communication primitive.

I. INTRODUCTION

With the Internet becoming the ubiquitous vehicle for all forms of communication, including information access and delivery, we argue that time has come to enhance the original end-to-end principle of the Internet [1]. The end-end principle is based on a host-to-host communication primitive. We enhance it to be a name-to-name principle in which communications take place among entities identified by names, without a static correspondence to their current location. We call such general framework Internames.

In Internames, names are used to identify everything: contents, users, devices, logical and physical points and services. The end points of every communication exchange are identified by names. Basic building blocks of Internames are: i) a name-based Application Programming Interface; ii) a separation of identifiers (names) and locators; iii) a powerful Name Resolution Service (NRS) that dynamically maps names to locators, as a function of time/location/context/service; iv) a built-in capacity of evolution, allowing a transparent migration from current location-oriented networks and the ability to include as particular cases current specific architectures and technologies.

It is well known that large parts of this picture have already been proposed, several times, by several researchers (e.g. [2][3][4][5][6][7]), meaning that the research community is ready for this leap, and that the zeitgeist or "spirit of the times" deems such vision mature enough to be feasible. Our contribution consists in adding another voice to this chorus, framing Information Centric Networking (ICN) architectures in a larger context. ICN as a concept [10][11], is attracting considerable interest (see papers [12][14][15] and the projects [16][17][18][19][20][21][22]). Most agree that the basic functions of an ICN infrastructure are to: i) address content, adopting an addressing scheme based on names, which do not include references to their location; ii) route a user request, which includes a destination content-name, toward the "closest" copy (original server, cache, end-users) of the content with such a name iii) deliver the content back to the requesting host. This is believed to offer several advantages [23][24][25][26][27]. We see however that several issues still need to be addressed: i) the need to devise credible migration paths from the communication paradigm of the current network infrastructure; ii) the complexity and scalability of the proposed naming and routing functionality; iii) the cumbersome support for push services; iv) several security and privacy concerns. It is necessary to justify that changing the current mode of operation in the network infrastructure is indeed worth the trouble. For example, some have suggested implementing ICN concepts only in end-devices, supported by existing network infrastructures and protocols [28].

In our opinion, there is a third way in-between a fully ubiquitous ICN as it is conceived e.g. in [12], hybrid ICN [13] and proposals such as [28]. Our aim is to design an architecture that enlarges the scope of ICN, extending its functionality beyond content retrieval, easing send-to-name and push services, and allowing the use of names to route data also in the return path. Unlike other proposals, Internames seeks to build upon and reuse existing protocol (e.g. HTTP, CCN, SIP, etc.) as a way of achieving gradual migration, and use Internames as the glue that interconnects different network realms. For instance, when communicating with an IP-infrastructure, by mapping names to locations and working through appropriate gateways we enable a name-oriented infrastructure to inter-operate with IP. Such a gateway could also serve as the bridge between API primitives of known protocols (e.g. HTTP GET or CCN GET) and the future name-based APIs of Internames.

While the concepts in the architectures being considered for ICN/CCN/NDN use names instead of locations to address a remote end-point, the source of the communication is still identified by the end-point and its current location (although, these architectures adapt to the receiver's mobility). But, we see a need to evolve from such a host-to-name communication to a name-to-name communication, where the source is also identified by a name. Also, we believe that names should be used to identify also users, devices and logical and physical points and services. The basic API offered to all applications should accept names as identifiers of requested contents or services. Then, an NRS would map names to network locations, as a function of time/location/context/service, reducing some of the problems of scalability, thanks to the identifier/locator split.

An NRS plays a key role in Internames, to enable the co-existence of multiple network domains, which we call *network realms*. Similarly to the role played by IP to unify different network technologies, Internames would unify ICN (and, possibly, different flavours of ICN), IP networks and non-IP networks. The NRS would reconcile and unify several network technologies, IP, cellular, sensor, IoT, ad-hoc, mobile, but also new ICN networks. The NRS would map a name not only to a network locator but also to the right protocol to be used to reach the current location of that name. In the same way unicast, multicast, broadcast, anycast communications would be a property bound to names, with the resolution service mapping names to requested services. In this sense, the NRS is more powerful than the current DNS, and extends the functionality that has been suggested in the Global Name Resolution Service (GNRS) by the MobilityFirst project [5]. Is such an NRS feasible? Time will tell. However, today's technologies allows in principle for implementing a logically centralized NRS coupled with localized instantiations of functionality, fully in line both with the recognized need of network abstractions (as theorized in Software Defined Networking approaches [8]) and implementations exploiting Network Function Virtualization [9] and Cloud principles.

What will Internames enable as a future network paradigm? We believe it will allow named-entities to be mobile and be connected to the network infrastructure anywhere, enable them to be reached by any of a number of basic communication primitives, allow communication to span networks with different technologies and allow for disconnected operation. Where it is feasible, we believe that having the appropriate transport layer primitives, communication would progress uninterrupted as a named entity moves from one interconnection point to another. Furthermore, with the ability to communicate between names, the communication path can be dynamically bound to any of a number of end-points, and the end-points themselves could change as needed.

II. INTERNAMES: A NAME TO NAME PRINCIPLE

Internames enables communications among *named-entities*. The difference from existing ICN approaches is that it is designed to use names not only within an ICN proper but for communicating over any network. Using names as the primary means for applications to access entities and a name-based routing and forwarding plane (e.g., name-based routers), Internames aims to do for content and services what IP did with IP addresses and routers: create a glue to interconnect networks, potentially of different technologies. In this vision, names provide access to content and service access points distributed on networks of any type, including public/private IPv4/v6 networks; public/private overlay/clean-slate ICN and IoT networks; Data Centers and Cloud; ad-hoc/mesh/cellular networks; DTNs; etc.. Different name spaces could be used to allow the architectural framework to adapt to the different needs of different contexts or the interpretation of the names could vary, based on the context. Names could be associated with additional meta-data information, e.g., description of the content, rights to use it; expiry date for the content, capabilities of devices etc.

To better motivate the name-to-name communication framework, we describe a use case developed in the framework of the "GreenICN" joint EU-Japan project [22]: that of Disaster Management, including prevention, detection and recovery from the effects of disaster on ICT infrastructures where every person, sensor or monitoring system may be a sender and/or receiver of information. When there is a single sender of information, to one or multiple recipients identified by a name, current CCN/NDN host-to-name communications are sufficient. However, in our use case, each sender may have different roles, personas and responsibilities (as an individual, as an authority). When that person wishes to send some information, e.g., related to a disaster, the initial communication could be viewed as coming from an authority (identified by a name) to a designated set of recipients (identified by another name). Thus, unlike CCN/NDN, the return path for communication need not follow the original path and can instead be addressed to the authority, i.e. to a name. The responses would be delivered to the original name that transmitted the initial message, even if the named entity moves from its original location. Moreover, the response could be delivered to all entities associated with that name (in a sense, it is the reverse of a traditional 'multicast', having the information flow from receivers to all senders). This has the potential to reduce the amount of state maintained in the network (e.g., in the Pending Interest Table in CCN/NDN enabled routers). In addition to performing name-based forwarding in both directions, network components such as the NRS, intermediate routers etc. could perform additional services such as sender/receiver name-based authorization, and facilitation of communication between a group of senders and receivers where each set is identified by a single name. Now, we describe the architecture and its components.

## A. Internames: Architecture and functional components

**Architectural drivers** – Internames is designed to also reuse current technologies and accommodate future technologies. Accordingly, the architecture:

- does not require the use of a fixed naming scheme, but is flexible in supporting current (e.g., DNS, CCN, EPC) and future naming schemes.

- unlike IP, does not specify a protocol data unit or layer but it can use existing and future name-oriented protocols, such as HTTP, CCN, SIP, DNS, etc.

- uses a Name Resolution Service to discover the forwarding rules that must be used by nodes in the network path. The Name Resolution Service is a backward compatible extension of the current DNS.

- exposes applications a name-oriented API, whose primitives allows fetching, pushing, publishing and subscribing to data through names.

- uses new name-based routers, and uses proxy (or gateway) technologies to bridge API operations among autonomous networks, at least as a temporary solution (waiting for a possible ubiquitous name-based protocol).

The end result is a unified framework with universal inter-operability, simplifying the current network architecture.

**Name-oriented API** – Internames exposes to applications a name-oriented API that enables *pull*, *publish* and *subscribe* to content identified by names. And to *push* data towards the communication interface (port) of an application (i.e., a service access point) that is identified by a name. The API provides also search primitives based on keywords and metadata.

**Namespace** - As shown in Figure 1 the architecture uses a *namespace*, where names are associated with entities. Entities may be digital data (content, information, etc.) or service access points through which an application can send or receive data (e.g., a TCP/UDP port). We refer to an entity associated with a name using the term **named-entity (NE)**.

**Name-realms** - The namespace is formed by *name-realms* that are disjoint containers of names, managed by different administrations. Name realms use local name schemes that may differ among them. A generic Internames name (e.g., n2n://nriA:Alice.com/cell) is a URI composed by a name-realm identifier (e.g. nriA) followed by an identifier that uses the local naming scheme (e.g. Alice.com/cell). For instance, a name-realm may be a set of names using a naming scheme as proposed by CCNx, or may be the set of current DNS names, or EPC flat identifiers.

**Network-realms** - A named-entity is dynamically or statically bound to one or more Network Attachment Points (NAPs), i.e. addressable network ports or interfaces, available possibly in different *network-realms*. A network-realm is an autonomous network using a local networking stack (e.g., IP, Ethernet, ICN, etc.), and whose routing scope is bounded to that network domain only. For instance, a network-realm may be the public Internet, an ICN, a Content Delivery Network, or a Data Centre/Cloud. Network-realms may be *nested*. A nested network-realm uses the networking services of the underlying realm to interconnect its nodes. Practically a nested network-realm is an overlay network and, for instance, it may be an autonomous CCN within the public Internet network-realm.

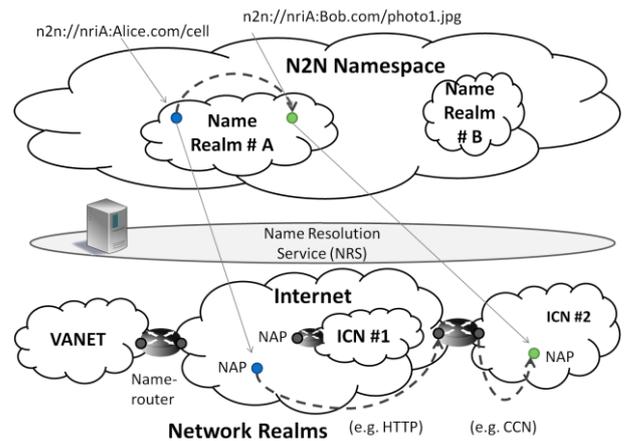

Figure 1 Architecture components

**Name-router -** Network-realms are interconnected to each other through a dedicated node called *name-router*. Different from IP or CCN routers that forward their own protocol data units, a name-router operates as a *protocol bridge* between the name-oriented protocols operating in each of the interconnected network-realms. For instance, as shown in Figure 1, a name-router could receive an HTTP request coming from a NAP hosted in the public Internet and "translate" it into a CCN request, when the destination named-entity is contained in an ICN network based on CCN technology.

**Service Descriptor (SD)** – Within a network-realm a named-entity can be accessed using a name-oriented protocol. By the term name-oriented protocol we refer to a communication facility used to pull, push, subscribe or publish data by name, e.g., HTTP (note that HTTP 2.0 will support a PUSH primitive), CCN, SIP, etc. The metadata that describes the name-oriented protocol operation to access a named entity is called a Service Descriptor. A Service Descriptor may match a single named-entity or a set of named-entities, e.g., whose names share a common prefix. For instance, in an IP realm the principal of cnn.com may use a service-descriptor that specifies how to relay HTTP requests to named-entities whose names start with cnn.com, to a given HTTP forward proxy. The service descriptor could be of the format: *<protocol, FCN, next-hop phy type, next-hop address>*. Here, FCN, is the **Forwarding Component Name,** the name that could be used in a CCN realm to perform name-based forwarding (explained later). An example of a Service Descriptor is as follows: Let us assume that the content is available in a CCN network realm. The Service Descriptor could look like: *<protocol=CCN/UDP, FCN = ccnx:/ccn.com/article.pdf, next-hop phy type = IP, next-hop address = 160.80.80.1>*, in which 160.80.80.1 is the IP address of the Name-router that is present at the entry point of the CCN realm and is therefore the next CCN router of the path towards the ccn.com repository.

**ORS -** A searchable database that we call **Object Resolution Service (ORS)** contains all names of the namespace and related metadata. Users searching for a content or a service access point would query the ORS (as in a current search engine) and get a name or a list of names and the associated metadata, if available. The ORS is not strictly necessary, and may be replaced by private search engines and could also be implemented in a distributed manner.

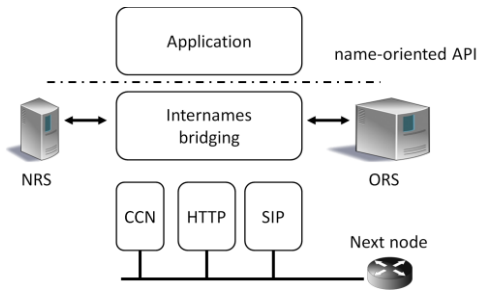

Figure 2 Node Functionality

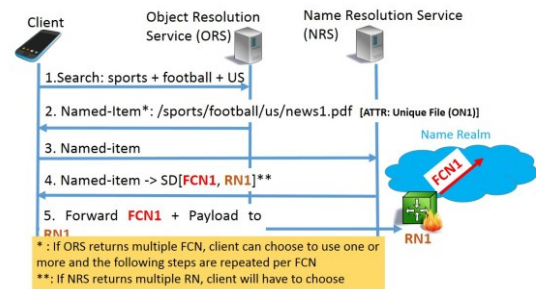

Figure 3 Message flow

**NRS** – The **Name Resolution Service (NRS)** is a powerful and flexible resolution service that maps a named-entity to a Service Descriptor. When a node (an application or a named-router) needs to forward a request towards a named-entity it uses an NRS to find a Service Descriptor to relay the request to the next-hop node. This may be the next reachable name-router of the path (i.e., to enter a network realm) or the final destination (e.g., IP address of the repository). Resolved service-descriptors may be cached. The NRS architecture may follow DNS and the interaction may be an extended version of the DNS protocol (e.g., starting from RFC6891). The resolution could be a function of time, context, location and requested services. For instance, in a disaster situation in which a section of the network is separated from the big Internet, names would be resolved to different locators in contrast to normal conditions. It would be performed transparently to users, who can continue using their applications as usual, as far as possible. In case there are multiple suitable Service Descriptors, the NRS could either decide to respond with only one of them or return multiple Service Descriptors along with attributes that better describe them.

**FCN** – The **Forwarding Component Name (FCN)** can be used to perform name based forwarding in a CCN realm. These names are in the form of a fully qualified domain name (FQDN, as it has been traditionally considered, possibly with a hierarchy). The FCN could be a self-signed name, a hierarchical name, or a hierarchical topic name in the case of topic based pub/sub. Therefore, in certain cases, there could be a one-to-one mapping between a named entity (NE) and an FCN (see Figure 3) or a many-to-one mapping between NE and FCN (in the case of pub/sub for a topic).

**Abstract Layering -** Figure 2 shows the functionality of a node. Applications use a name-oriented API to interact with a named-entity. For instance, for retrieving a content item by using its name or for pushing a talkspurt of speech to a mobile phone application, the API provides access to a named-entity. Below the API, a bridging function serves the primitive call by relaying the API request to the next node on the path according to named-oriented means (CCN, HTTP, SIP) indicated in the Service Descriptor resolved by the NRS.

*B. ICN overlay deployment within the Internames framework*

This subsection describes a possible IP overlay deployment of an ICN technology within the Internames framework. Let's assume that the network is formed by ICN network-realms (NRs). A NR is a closed ICN serving named-entities. Applications running on the user terminal could be in the Internet NR and reach the access name-router of the ICN NRs through simple HTTP. The terminal queries the NRS to discover that the HTTP protocol is to be used to talk with the Access name-router. The terminal may additionally use an ORS to obtain a list of named-objects that match some search criteria (e.g., keyword, metadata, etc.).

Different from other ICN architectures that suffer from routing and security issues, the separation of the network into "isolated" realms alleviate these problems. The routing plane of an NR only manages name-prefixes of the offered named-entities and does not import any other route from named-entities of other NRs, thereby reducing the number of routing entries of the routing plane. An ICN NR is an autonomous content network wherein the owner could autonomously decide forwarding, caching, security, and other content-based strategies. An NR may be a CCN network of a private company whose functionalities are specialized to serve a particular class of applications, a CDN network, or a mobile network that offer content-based services to its customers, etc. An NR can even be engineered to support one specific application, e.g. video delivery, health care, etc. It is also possible to perform access control on dedicated Name-router limiting security issues, such as injection of fake contents by untrusted users or CCN PIT flooding. In case the content-oriented protocol used by the terminal to contact the Name-router differs from the one used inside the NR, the Name-router could behave as a proxy/firewall between the NR and the outside network. E.g., the terminal uses HTTP and CCN is used within the NR.

Figure 3, shows an example of a message exchange. Let us assume that the client sends a list of key words to the ORS, which in turn finds one or more Objects that satisfy this query. The ORS returns the corresponding Named-entity along with meta-data. The client will forward the returned *Named-entity* to the NRS in order to obtain the SD. The NRS, in this example, returns the SD *<protocol=CCN/UDP, FCN = FCN1, next-hop phy type = IP, next-hop address = RN1>*. The client uses the returned SD to send the request to the next hop (RN1). RN1 will further send the request in the CCN realm using the FCN1 as the name to perform name based routing.

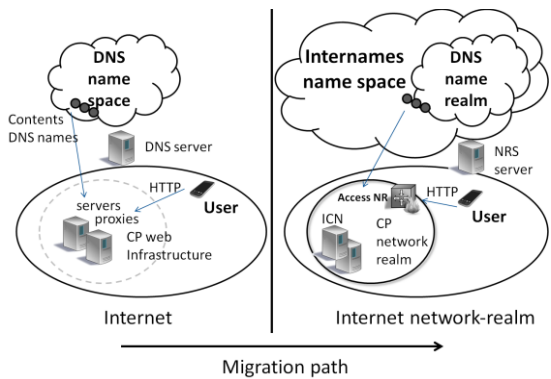

**Figure 4 Migration path (left: before migration; right: after migration)**

### III. MIGRATION PATH AND EXAMPLES

Internames inherently supports a smooth migration from current networks. Indeed, name and network realms can be progressively deployed. The NRS technology can easily be compatible with DNS and the DNS namespace can be considered as the "first" name-realm of Internames namespace, as much the current Internet can be considered as the first network-realm. Production IP-based services are transparently merged in the Internames architecture. Operators do not need to switch from IP to ICN. In what follows we discuss possible steps of network actors (content providers, network carriers, and users) to migrate IP/DNS services to Internames services.

**Content-provider** – Consider a content provider (CP) whose business is storing and distributing contents by using its own distribution network. For instance, current companies that offer CDN services. Today, a CP delivers contents to users by using an infrastructure formed by web, proxies and DNS servers (Figure 4-left). Contents have names belonging to the DNS namespace. Users access contents located on servers and/or proxies through HTTP means. As shown in Figure 4-right, a first step of the migration consists of replacing the authoritative DNS server of the provider content with a NRS server. This does not perturb production services, since NRS is backward compatible with DNS. A second step is to deploy a private network-realm, connected to the Internet through a name-router. The connectivity of such network-realm could be an overlay IP network; i.e. the network-realm will be nested in the Internet one. The networking technology used within the network-realm (e.g. ICN) may be optimized for the distribution of the specific provider contents (e.g. videos). The final technical step consists in updating the NRS with proper Service Descriptors, which uses the IP address of the name-router as next-hop and indicates HTTP as name-oriented protocol. Now contents that were distributed in the Internet network-realm e.g. though HTTP means are migrated on the new network-realm.

**ISP** – Consider an Internet Service Provider (ISP) whose business is the transport of data from/to user premises or customer networks. For instance companies providing fixed and wireless internet access, or carriers of the Internet Tier-1. ISPs may deploy new network-realms (possibly nested in their actual IP infrastructure) either to build name-based transit networks, interconnecting newly deployed name-realms of content-providers, or to proxy name-based services that are currently available on the Web (Internet Realm). The transit service may use the same technology of the interconnected realms, so avoiding CP access routers to perform costly protocol bridging functions. Proxy services may be useful to reduce inter-domain traffic, with an ISP realm implementing an ICN or HTTP proxies hierarchy.

**Users** –End-users can continue to use IP and DNS based services as NRS would be DNS compatible; access routers of newly deployed network-realms would support the proxy of current name-based protocols such as HTTP. In this case, end-user devices belong to the Internet realm. In an evolutionary scenario end user-devices can have interfaces connected (directly or via tunnels) to newly deployed network realms.

### IV. HOW INTERNAMES ADDRESSES DISADVANTAGES AND ADVANTAGES OF CURRENT ICN

In this section we discuss how known cons and pros (e.g. [23]) of ICN are reflected in Internames.

#### A. Disadvantages

**Migration path -** Transition to a whole new kind of router scares operators and manufactures and investors. With Internames evolution would be gradual and mostly concentrated in network edges and hosts. The resolution server can be initially the current DNS, which can gradually and smoothly evolve in a more powerful and dynamic and localized version of itself.

**Routing scalability -** A full ICN implies very large routing tables and high frequency of updates. Internames envisages the presence of multiple network realms including pure ICN realms, pure IP realms or a hybrid between the two. This allows various levels of separation that could facilitate efficient management of routing tables. Moreover, routers in the network need not have the same capability and can be grouped based on their functionality and capability into different NRs. Furthermore, Internames relies on name-to-name communication also for the return path, and therefore does not need additional states in the router in terms of PIT.

**Cumbersome support for push services -** ICN comes in several flavours. Some do not support push services, while in architectures such as [17] procedures are required that implies different and heterogeneous network entities that unfavourably compare to the simplicity and homogeneity of IP routers. In Internames push services would be supported just as of today, once a name is mapped to a network locator.

**Security -** In ICN, users (in some instances even end-users) should be allowed to update the routing plane to make their content reachable; the stateful forwarding of Pending Interest Tables [12] could be flooded with fake instructions. The resolution server of Internames, and thus routing plane information, would be under the control of trusted organizations and not of end-users. Additionally, the name-router at the entry points of network realms provides another degree of security, as explained above.

#### B. Advantages

Internames allows providing most of the advantages promised by ICN. It can perform content-based routing, if the NRS has sufficient information; it can provide off-path and in-path caching, not only in ICN routers, but also in IP routers using mechanisms similar to that proposed in [29]. Internames, as ICN, can facilitate mobility and can offer all the advantages of content-based operations, including content-oriented security, content-oriented access control, content-oriented QoS differentiation (and possibly pricing). Since the NRS is aware of the transferred content, Internames allows to better control information and related revenue flows. Finally, name to name communications and bidirectional, two-way links between contents could become a reality as envisioned in works such as [30]: "In a network with two-way links each node knows what other nodes are linked to it.. and preserve context".

## V. Conclusions and Mobility-related issues

Implementing a full ubiquitous ICN would mean to radically change the way the Internet works today, including changing routers, applications, and networking stacks. On the other side, if we give up implementing core ICN functionality within the network, such as routing-by-name and forwarding-by-name, then ICN would collapse either to a CDN internetwork or in application-layer ICN, exploiting HTTP, or evolution of HTTP, without network-layer support.

We propose a third way, which avoids the pitfalls of scalability and deployment issues, by confining ICN operation in section of the networks, leaving (initially) untouched its core. We believe that ICN could profitable be used in mobile access and wireless networks, and in ad hoc environments without infrastructure support, such as isolated sections of the networks and IoT scenarios. In these environments, scalability and deployment issues are of less importance than in the Internet at large. In addition, as noted in recent papers (e.g. [28]), in-network caching and performance gains brought about by ICN within the network are not as compelling as initially foreseen by ICN advocates. Instead, expected benefits of ICN such as mobility support, ability to work in un-infrastructured modes, peer-to-peer communications support, content-oriented security model, content-oriented access control and QoS differentiation and network awareness of transferred content are very attractive and indeed attainable in the mentioned environments. This architecture, and a suitably designed NRS, would easily support mobility, empowering advanced mobility functionality for every kind of network. The challenge is in designing and implementing the NRS to be scalable and to have high performance. We expect to learn from past efforts in evolving the DNS and recent efforts in the MobilityFirst project, as we pursue the implementation of our proposed architecture.